\documentclass[aps,prl,twocolumn,groupedaddress]{revtex4}
\usepackage{amssymb,amsmath,bm}
\usepackage{epsfig}
\usepackage{graphicx,psfrag}

\begin{document}

\title{The Mass of a Spin Vortex in a Bose-Einstein Condensate}
\author{Ari M. Turner$^1$}
 \affiliation{$^1$Department of Physics,
University of California, Berkeley, California 94720, USA}
\date{\today}

\begin{abstract}
In contrast to charge vortices in a superfluid,
spin vortices in a ferromagnetic condensate move inertially (if the condensate
has zero magnetization along an axis). The mass of spin vortices
depends on the spin-dependent interactions, and can be measured as a part
of experiments on how spin vortices orbit one another. 
For Rb$^{87}$ in a $1\ \mu$m thick trap, $m_v\sim10^{-21}$kg.
\end{abstract}

\maketitle
\date{\today}

Vortices, with their long-life and concentration of energy, 
often provoke comparison to particles. But does their \emph{motion}
fit the analogy?
Ordinary vortices in a fluid or superfluid do not move inertially, as 
particles
do, because their motion is Magnus-force dominated. For example, in the
absence of a force, they do not move. When a force is applied,
they move perpendicular to it, a situation which is 
described by first order
differential equations\cite{batchelor}.
In fact, the motion of a pair of vortices is a miniature version of
how Descartes\cite{cajori} explained the motion of planets, with
the sun causing the ether to whirl around,
dragging the planets at the same speed.  
On the other hand, spinor superfluids\cite{ho98,ohmi,lightandspinreview} 
made out of laser-cooled atoms can have
spin-current vortices.
These vortices will be argued to obey Newton's laws at low speeds; 
in particular
they have a mass, which determines their resistance to being accelerated.

A mass can also be defined
for a vortex in a superconductor or a superfluid and it may play
a role in determining
the oscillation frequency of vortex lattices and the tunneling rate of 
vortices\cite{kopnin}.  
Observing the mass for such vortices is much more subtle than for
spin vortices, though, because the vortex inertia is nearly overcome
by the Magnus force.  
The picture of a moving spin vortex described here is inspired partly
by the motion of vortex rings\cite{donnellyvortices} and Ref. \cite{jonesroberts}'s
study of that motion based on the Gross-Pitaevskii equations.


We will focus on the case of spin 1 ferromagnetic atoms in a two-dimensional
condensate.  A magnetic field is applied along $z$ to stabilize
the spin vortices, via the quadratic Zeeman effect. (We will take the condensate
to be in the $xy$ plane for definiteness, but the spin and spatial
coordinates can be chosen independently.)
The ground states of these atoms
have the form 
\begin{equation}
\psi=e^{i\theta_C+i\theta_S S_z}\left(\begin{array}{c}\sqrt{n_1}\\ \sqrt{n_0}\\ \sqrt{n_{-1}}\end{array}\right)=
\left(\begin{array}{c}e^{i(\theta_C+\theta_S)}\sqrt{n_1}\\ e^{i\theta_C}\sqrt{n_0}\\
e^{i(\theta_C-\theta_S)}\sqrt{n_{-1}}\end{array}\right).
\label{eq:moreorless}
\end{equation}
The values $n_1,n_0,n_{-1}$ correspond
to the optimal proportions of the three spin states (the total density 
is $n=n_1+n_0+n_{-1}$), and are fixed.
If the magnetization along the $z$-axis, $M_z=n_1-n_{-1}$, vanishes,
then the
spin is
in the $xy$-plane, as is
preferred by the quadratic Zeeman effect\cite{lightandspinreview}.
The overall phase of the
wave function (which is not observable) 
is $\theta_C$, and $-\theta_S$ is the azimuthal angle
of the spin. (To see this, calculate $<S_x>$ and $<S_y>$ for this
state.) The latter angle can be measured by scattering polarized light off
the condensate.

\begin{figure}
\psfrag{0}{0}
\psfrag{+}{+1}
\psfrag{-1}{-1}
\includegraphics[width=.45\textwidth]{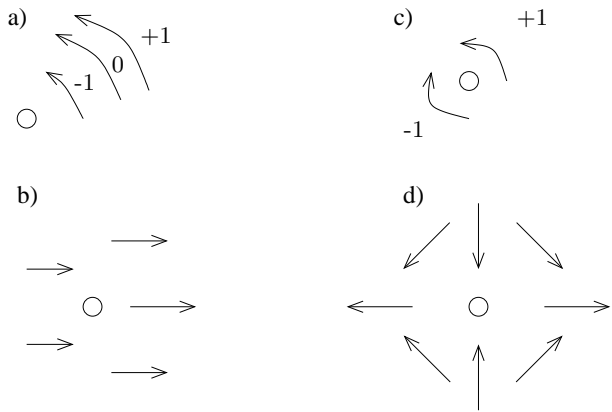}
\caption{\label{fig:starfish} Charge and Spin Vortices of strength $2\pi$. 
a), c) The currents of the spin states
around charge and spin vortices. b), d) The spin textures around charge and
spin vortices. }
\end{figure}

There are two types of
vortices in a spinor condensate in a magnetic field: 
a charge vortex, described by $\theta_C=\pm\phi,\theta_S=0$
and a spin vortex described by $\theta_C=0,\theta_S=\pm\phi$ (and
observed in a rubidium-87 condensate\cite{tiedye}). Here, $\phi$
is the azimuthal angle centered on the vortex core.
Fig. \ref{fig:starfish} illustrates these vortices:
\begin{equation}
\psi_C=\left(\begin{array}{c}e^{i\phi}\sqrt{n_1}\\ e^{i\phi}\sqrt{n_0}\\ e^{i\phi}\sqrt{n_{-1}}\end{array}\right)\ \ \ \ \psi_S=
\left(\begin{array}{c}e^{i\phi}\sqrt{n_1}\\ \sqrt{n_0}\\
e^{-i\phi}\sqrt{n_{-1}}\end{array}\right).
\label{eq:vortices}
\end{equation}
The spin texture is uniform around a charge vortex (except near
the core); the spin direction
rotates by $360^{\circ}$ clockwise or counterclockwise
in a spin vortex.  See Fig. \ref{fig:starfish}. 

A spinor condensate is somewhat similar to a mixture of several species
of atoms.  From this perspective, a spin vortex is a bound state of
two opposite vortices in two different components and a charge vortex is a bound
state of three vortices.  

The velocity fields
in the components of a spinor condensate 
are related:
\begin{equation}
\bm{u}_{1}=\bm{u}_C+\bm{u}_S;\ \ \ \bm{u}_{1}=\bm{u}_C;\ \ \ \bm{u}_{-1}=\bm{u}_C-\bm{u}_S;
\label{eq:coordination}
\end{equation}
The flow in the middle component (for spin 1) is the mean of the flows
in the other two components.  As a simple example, the atoms
near a charge vortex flow with velocity $\bm{u}_m=\frac{\hbar}{M}\nabla\theta_m$
\cite{donnellyvortices}. They all move
in the same direction (see Fig. \ref{fig:starfish}a),
meaning that there is a net transport of mass.  On the other hand, the
atoms near a spin vortex do not carry any mass if the condensate
has zero magnetization $M_z=n_1-n_{-1}$, because the atoms of spin $\pm 1$ move at equal
and opposite speeds.  However, there is a net transport of spin, since
the amount of angular momentum carried counterclockwise
depends on the current of $m=1$ atoms
moving counterclockwise and of $m=-1$ atoms moving clockwise.
Vortices
can be classified by the amounts $\theta_C$ and $\theta_S$ wind by,
$Q_C$ and
$Q_S$ respectively.
(In general, the charge and spin currents are given by
$J_S=n\bm{u}_C+M_z\bm{u}_S$
and $J_S=q_z\bm{u}_{S}+M_z\bm{u}_C$ where
$q_z=\psi^{\dagger}S_z^2\psi$. If $M_z=0$, charge vortices have only charge
currents and spin vortices have just spin currents.)

Now the force on a vortex moving at speed $\bm{v}$ and
tossed about by a flow of charge and spin
is given by 
$\bm{F}=-\hbar\bm{\hat{z}}\times[\sum n_mQ_m(\bm{u}_m-\bm{v})]$.
The term for a given value of $m$ describes
the lift force on the
vortex in that component: according to Bernoulli's principle
higher velocities correspond
to lower pressures, so the vortex moves to the side where its velocity field
is pointing in the same direction as the relative velocity of the fluid.
For a spin vortex of charge $Q_S$ the total
force is
\begin{equation}
\bm{\dot{p}}=\bm{F}_S-Q_S\bm{v}\times \hbar M_z\bm{\hat{z}},
\label{eq:meteor}
\end{equation}
where $\bm{F}_S$ is a force produced by the background spin current and
the second term is the Magnus force, or lift, responding to the vortex's
own motion.
For a charge vortex of charge $Q_C$, the force (assuming $M_z=0$) is
\begin{equation}
\bm{\dot{p}}=\bm{F}_C-Q_C\bm{v}\times \hbar n\bm{\hat{z}},
\label{eq:restful}
\end{equation}
where $\bm{F}_C$ is produced by charge current.

\emph{Idiosyncracies of Vortex Motion:}
Let us compare the motion
of charge and spin vortices.
A charge vortex in stationary fluid cannot
drift along a straight line, because the lift force would push
it sideways.  Furthermore, the equations of motion take the form
of first order differential equations for charge vortices when
the inertial term $\bm{\dot{p}}$ in Eq. (\ref{eq:meteor}) is assumed small:
\begin{equation}
\frac{d\mathbf{r}}{dt}=\frac{1}{n\hbar Q_C}\bm{\hat{z}}\times \bm{F}_C;
\label{eq:dizzying}
\end{equation}
motion is perpendicular to applied forces.
The vortex velocity required by Eq. (\ref{eq:dizzying}) turns
out to equal the background flow speed, in accordance with Descartes's
conception of planetary motion.
\emph{The motion of charge
vortices is determined once their initial positions are given}.

In contrast, a spin vortex behaves in a Newtonian way, 
as long as the condensate has
zero magnetization. 
In the absence of spin current, Eq. (\ref{eq:meteor}) implies
``Newton's first law of spin vortices":
\emph{a spin vortex in a charge current
can move at any constant speed}. 
There is no lift to push the vortex off course
in an unmagnetized condensate because the component
vortices $\psi$ rotate
in opposite directions (see Fig. \ref{fig:starfish}c). 
Now if there is a spin current
the lift-forces on the component vortices
from the counterpropagating flows \emph{add} to produce
a nonzero $\bm{F}_S$. The solution to
$\bm{\dot{p}}=0$, 
$\bm{v}=\frac{\bm{F}_S}{\hbar Q_S M_z}$, does 
not make sense if $M_z=0$. 
Therefore the inertial term cannot be neglected
and ``Newton's second law of spin vortices" results: a \emph{spin-vortex in
a spin current must accelerate}, at a rate proportional to $\bm{F}_S$.
The spin force has an electrostatic form:
\begin{equation}
\bm{F}_{S12} =\frac{\hbar^2 q_z}{M}
\frac{Q_{S1}Q_{S 2}\bm{\hat{r}}_{12}}{r_{12}}.
\label{eq:electricfield}
\end{equation}
One can introduce the
vortex mass $m_v$ by assuming that 
\begin{equation}
\bm{p}=m_v\bm{\dot{r}}
\label{eq:nicetomeetyou}
\end{equation}
at least at low
speeds.  
Hence, the equation of motion prescribes the \emph{acceleration}:
\begin{equation}
m_v\frac{d^2\bm{r}}{dt^2}=\bm{F}_S\label{eq:Newton}
\end{equation}
These Newton's laws describe spin vortices in condensates of any
spin, as long as $M_z=0$\footnote{More
precisely, the cross-stiffness $K_{CS}$ must equal zero.
When the Gross-Pitaevskii
equations apply, $K_{CS}\propto M_z$. Also,
$K_{CS}=0$ for any time-reversal symmetric state.}.


\emph{Phase Space:}
The phase space of $N$ vortices seems likely to be 
$2N$ dimensional, given that
the basic equation for the evolution of a spin-1 condensate
is the following first-order differential equation:
\begin{equation}
i\hbar\frac{\partial\psi}{\partial t}=-\frac{\hbar^2\nabla^2}{2M}\psi+\frac{\partial\mathcal{V}}{\partial\psi^{\dagger}}
\label{eq:notanegg}
\end{equation}
Here $\mathcal{V}$ is the potential energy of the atoms. 
Only the \emph{spatial} coordinates of the vortices seem essential for
parameterizing the superfluid wave function.

For charge vortices, 
the phase space \emph{is} $2N$ dimensional. The motion
is even described by Hamilton's equations\cite{lin}, with
the $y$ coordinate
of each vortex acting as the momentum conjugate to $x$!
(There are surprising
consequence for the thermodynamics of vortices\cite{onsagerthermo,montgomery}.)
The precise conjugacy relation reads
\begin{equation}
\{x,y\}=\pm\frac{1}{nh},
\label{eq:descartes}
\end{equation}
where the sign depends on the direction of circulation. 

However, Newton's second law implies 
that $N$ \emph{spin} vortices have
a $4N$ 
dimensional phase space.
For ordinary objects, the momenta
exist only in an abstract space:
a photograph of a ball in the air does not reveal its destination.  
The future of a spin vortex will also depend on
which way it is moving at a given time,
so Eq. (\ref{eq:notanegg}) implies that something
about the wave function which is
\emph{observable in a single photograph} can be used
to \emph{deduce} the vortex's velocity. 

\begin{figure}
\psfrag{psi+}{Vortex in $\psi_{+1}$}
\psfrag{psi-}{Vortex in $\psi_{-1}$}
\psfrag{v}{$\bm{v_x}$}
\psfrag{Dy}{$D_y$}
\includegraphics[width=.45\textwidth]{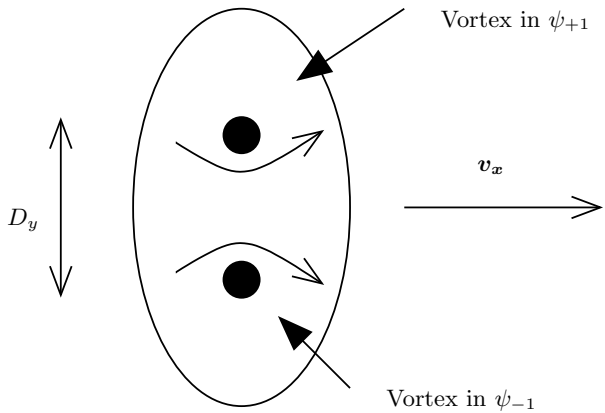}
\caption{\label{fig:pips} 
Seeing the momentum of a spin vortex. The phases of $\psi_{\pm1}$ increase
in the directions of the arrows. The component
vortices are pulled apart by the lift force to a distance $D_y\propto p_x$.  
The stretching
is limited by the phase-locking produced by the spin-dependent interaction.}
\end{figure}

To guess the tell-tale trait, note
that the lift forces on the component vortices of a moving
spin vortex pull them in opposite directions, but
they are restrained from drifting apart completely (see below).
Thus, the stretching of the
spin vortex increases with its speed
 (see Fig. \ref{fig:pips}).
The momentum $\bm{p}$ can be discerned
from a snapshot if it is proportional to the
stretching:
\begin{equation}
\bm{p}=-h n_1\bm{\hat{z}}\times\bm{D}
\label{eq:redwood}
\end{equation}
where $\bm{D}$ points between the component vortices.
This relation is consistent with the
canonical mechanics of the component vortices\cite{ashvin}.
According to
Eq. (\ref{eq:descartes}) 
$\{x_\pm,y_\pm\}=\pm\frac{1}{n_1h}$
where $(x_\pm,y_\pm)$ is the location of the component
vortex with circulation $\pm 2\pi$.
Now the center of mass coordinates $x$ and $y$ of the
spin vortices have the ordinary
Poisson bracket of zero, but $x$ is
conjugate to $p_x=n_1 h(y_+-y_-)$:
\begin{equation}
 \{x,p_x\}=1.
\label{eq:jackandjill}
\end{equation}

Eq. (\ref{eq:redwood}) is correct for a dipole of vortices
in a single-component superfluid also, but the velocity
of the dipole increases as its momentum (the distance between the vortices)
decreases!
This unusual fact can be traced back to the logarithmic interaction
energy $E\propto \ln D$.  Spin vortices behave
more normally because the confinement energy
is quadratic, so
$v=\frac{dE}{dp}$ increases linearly with $p$.

Eq. (\ref{eq:redwood}) holds when the phase-only 
approximation applies\cite{future}.
The general formula for
the momentum is 
$\bm{p}=-\bm{\hat{z}}\times\iint d^2\bm{u}\ \bm{r}\ \mathrm{curl}\ \bm{J}$,
where $\bm{J}$ is the mass current (see
\cite{saffman}).  In the phase-only approximation, $\mathrm{curl}\ \bm{J}$
is a dipole of delta-functions, leading to Eq. (\ref{eq:redwood}).

\emph{Confinement and the Vortex Mass:}
Spinor condensates can have special symmetries that
ordinary mixtures (e.g., of two different atoms) cannot have\cite{inert}.
But the higher symmetry is not the only special thing about 
spinor condensates.  Unlike in a mixture, the atomic states can turn into one another in
a spinor condensate, as long
as the angular momentum does not change: $1+-1\leftrightarrow 0+0$.
  This 
 ``coherent spin-flipping" process is
contained in the potential energy operator,
$\mathcal{V}=
\frac{1}{2}\alpha_{m_1m_2}:|\psi_{m_1}|^2
|\psi_{m_2}|^2:+2\beta\Re(\psi_1^{\dagger}\psi_{-1}^\dagger\psi_0^2)+q\psi^{\dagger} S_z^2\psi-\mu|\psi|^2$.

Coherent spin-flipping locks the phases of
the spinor components together\cite{pudynamics,chang1} 
and this keeps the two parts of a spin vortex bound together.
A nonzero phase $\lambda=(\theta_1+\theta_{-1}-2\theta_0)$ costs
energy
\begin{equation}
\mathcal{E}_{flip}=2\beta n_0\sqrt{n_1n_{-1}}\cos\lambda,
\label{eq:turnturnturn}
\end{equation}
where $\beta<0$ is the spin-dependent interaction parameter.
Since the energy
cannot depend on a phase unless particle numbers are not conserved,
it makes sense that this comes from the
spin-flipping reactions.

This ``coherent chemistry" interpretation of Eq. (\ref{eq:turnturnturn}) 
is complemented by an argument which views the atoms
as classical magnets: the
magnetization in the $xy$-plane depends on the relative phases
between the spin components and decreases as 
$\lambda$ increases. This
goes against the ferromagnetic propensities of the atoms.

Now
the \emph{potential energy} of the stretched vortex
\begin{equation}
E\sim|\beta| n^2 D^2
\end{equation}
can be reinterpreted
as kinetic energy on account of Eq. (\ref{eq:redwood}).  
At low speeds, $E=\frac{p^2}{2m_v}$, so
\begin{equation}
m_v\sim\frac{\hbar^2}{\beta} \sim M\frac{w}{\Delta a}.
\label{eq:smash}
\end{equation}
The second expression is obtained by relating  $\beta$
to the width of the condensate $w$ and
the scattering-length difference $\Delta a$ 
\cite{chang1}.
In this case, $m_v$ is of the same
same order as the mass of the atoms in the vortex core, $nl_m^2M$,
where $l_m$ is the magnetic healing length.
For rubidium in a $1\mu m$ wide trap with $\Delta a\sim 1$\AA,
$m_v\sim10^{-21}$ kg.

\emph{Measuring the Mass:}
Now consider the consequences of spin vortices' inertial motion.
Spin
vortices of opposite signs orbit around one another whereas opposite
charge vortices
push each other along parallel lines\cite{batchelor}.
As for planets, 
the orbits have different shapes depending on the initial  
momenta of the vortices. Specifically, when the two vortices
move on a circle, attracting each other according to
a $\frac{1}{r}$ force law, the period 
is proportional to its radius. Balancing the
attraction Eq. (\ref{eq:electricfield}) against the centrifugal force
$\dot{p}=p\omega$,
gives
\begin{equation}
vp(v)=\frac{2\pi n_1\hbar^2}{M}
\label{eq:kepler}
\end{equation}
where $v$ is the speed of both vortices.  This remains true when $p(v)$
is nonlinear.
The left hand side is an increasing function of $v$, so 
$v=v_{circ}$ is determined. This speed is
independent of the orbit radius.
 
Measuring $v_{circ}$ will give an \emph{estimate} for the vortex mass 
appearing in Eq. (\ref{eq:nicetomeetyou}).  However,
the linear relation for $p(v)$ is not reliable
at the speed $v_{circ}$ because
it is on the order of the speed of spin-waves in the condensate.

A more accurate way to measure the mass of spin vortices is to
observe their motion when the magnetization is not zero, but is small.
Then the lift force due to the vortex's
motion 
looks like the Lorentz force from a small magnetic
field, $\bm{B_A}=-M_z\hbar\bm{\hat{z}}$.
A single vortex will therefore follow cyclotron orbits with the period
\begin{equation}
\tau=\frac{m_v}{\hbar M_z},
\label{eq:cyclotron}
\end{equation}
If the magnetization is 5\% and the other parameters of
the condensate are those given above, this
period comes out to be $.3\ \mathrm{sec}$. 

\emph{Limiting Velocity:} A final idea for an experiment is to study the motion
of a rapidly moving vortex.  Such a measurement 
allows hypothetical inhabitants of the superfluid to measure the velocity of
their ``ether."  Spin vortices can move inertially
only up to a certain velocity relative to the
condensate.

Imagine pushing a spin vortex, starting from rest.  After a certain
amount of acceleration, the vortex may become unstable, so that
all the additional energy goes into producing spin waves.
Alternatively, the vortex may remain stable, and
absorb all the energy.  The
energy goes into stretching the components
apart until the vortex becomes needle-shaped.
The velocity is bounded in this case as well.
$E$ is proportional to the vortex-length $D\propto p$
and a linear dispersion implies a finite velocity.

\begin{figure}
\includegraphics[width=.4\textwidth]{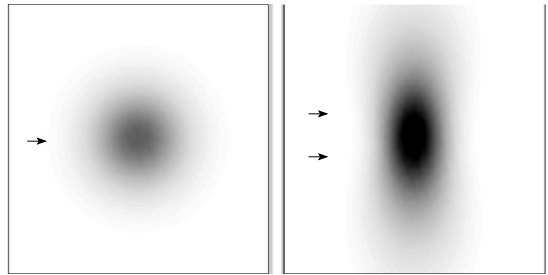}
\caption{\label{fig:marshmallow}
Computer solution for vortex cores moving at speeds
$v=0,.45\sqrt{\frac{\mu}{M}}$, with
rotationally symmetric interactions.  The darker regions have the higher
energy densities. The arrows indicate where $\psi_1$ and $\psi_{-1}$
vanish.
Note that the energy density extends past these zeros, where the
phase mismatch ends.  This energy
must come from varying magnetization, which
eventually causes spin wave emission.}
\end{figure}

Spin-wave dissipation seems to be the fate of a vortex in a condensate where
the atomic interactions in the Hamiltonian are rotationally symmetric
(of form $\frac{1}{2}\alpha (\psi^{\dagger}\psi)^2
+\frac{1}{2}\beta(\psi^{\dagger}\mathbf{S}\psi)^2$).
This conclusion is based mainly on numerical solutions of 
the Gross-Pitaevskii equations for steadily moving
vortices with $\beta=-.3\alpha, q=.5\mu$.
(see Fig. 
\ref{fig:marshmallow}). The computer
did not find solutions past 
$v_c=.65\sqrt{\frac{\mu}{M}}$, which is close
to the spin wave speed.
 
Experimental conditions can be adjusted so that the vortex does
not radiate sound.
The symmetry
of the condensate has to be broken further, 
by displacing the traps for the three
$S_z$ states into parallel planes.  
They still have
to overlap some to allow for interspecies conversion.
This trap set-up maintains the
 distribution of the atoms among the spin states
more rigidly\footnote{Separating the different spin
species increases the energy cost for changing the proportions of
atoms in the states from $q\sim\beta n$ (the spin-dependent interaction
energy) to $\alpha n$.
With the clouds together, the atoms can all
move into the $+1$ state without changing the $\alpha$
term.  But the density of the atoms increases when
they all go into the same spin state if the traps are displaced.}.
The component vortices can now separate arbitrarily far
without any instability. 

Displaced clouds also allow one to
use the component-vortex picture more rigorously\cite{future}.  When
$|\beta|\ll\alpha_{min}$ ($\alpha_{min}$ is the smallest
eigenvalue of $\alpha_{m_1m_2}$), the phase-only approximation
$|\psi_m|=const.$ applies. 
One can then derive the $p(D)$ relation Eq. (\ref{eq:redwood})
and show that the speed of an infinitely stretched vortex is 
$\frac{4n_0}{\pi}\sqrt{\frac{|\beta|}{nM}}$.

%

To summarize, each charge-vortex has only two spatial degrees of
freedom because the lift force overcomes the inertia. In
contrast spin
vortices in an unmagnetized condensate
behave like classical particles because they are made up of
oppositely rotating vortices whose total Magnus force cancels. 
The internal stretching between these components give rise to
the mass.  Spin vortices have four degrees of freedom: the
center of mass coordinates and the momenta
which are proportional to
the distortion of the vortex core.  The vortices behave like
classical particles at low speeds, but betray their composite origin
when accelerated sufficiently.

\emph{Acknowledgments} I have very much enjoyed conversations with Ryan Barnett,
Eugene Demler, Markus Greiner, and
Ashvin Vishwanath.  Markus Greiner started me on this research
by asking how to start a vortex moving in a spinor condensate.  
This research was supported by ARO Grant No. W911NF-07-1-0576.

\end{document}